\documentstyle[12pt,epsfig]{article}
\begin{document}
\centerline{\Large\bf Fine structure constant in the spacetime
of a cosmic string}
\vspace*{0.050truein}
\centerline{Forough Nasseri\footnote{Email: nasseri@fastmail.fm}}
\centerline{\it Physics Department, Sabzevar University of
Tarbiat Moallem, P.O.Box 397, Sabzevar, Iran}
\centerline{\it Khayyam Planetarium, P.O.Box 769, Neishabour, Iran}
\begin{center}
(\today)
\end{center}

\begin{abstract}
We calculate the fine structure constant in the spacetime of a cosmic
string. In the presence of a cosmic string the value of the fine
structure constant reduces. We also discuss on numerical results.
\end{abstract}

The gravitational properties of cosmic strings are strikingly different
from those of non-relativistic linear distributions of matter. To explain
the origin of the difference, we note that for a static matter
distribution with energy-momentum tensor,
\begin{equation}
\label{1}
T^{\mu}_{\nu}={\rm diag} \left( \rho, -p_1, -p_2, -p_3 \right),
\end{equation}
the Newtonian limit of the Einstein equations become
\begin{equation}
\label{2}
\nabla^2\Phi=4 \pi G \left( \rho+p_1+p_2+p_3 \right),
\end{equation}
where $\Phi$ is the gravitational potential. For non-relativistic
matter, $p_i \ll \rho$ and $\nabla^2 \Phi = 4 \pi G \rho$.
Strings, on the other hand, have a large longitudinal tension. For a
straight string parallel to the z-axis, $p_3=-\rho$, with $p_1$
and $p_2$ vanish when averaged over the string cross-section. Hence,
the right-hand side of Eq.(\ref{2}) vanishes, suggesting that straight
strings produce no gravitational forece on surrounding matter.
This conclusion is confirmed by a full general-relativistic
analysis. Another feature distinguishing cosmic strings from more
familiar sources is their relativistic motion. As a result, oscillating
loops of string can be strong emitters of gravitational radiation.

A gravitating string is described by the combined system of Einstein,
Higgs and guage field equations. The problem of solving these coupled
equations is formidable and no exact solutions have been found to date.
Fortunately, for most cosmological applications the problem can be made
tractable by adopting two major simplifications. First, assuming that
the string thickness is much smaller than all other relevant dimensions,
the string can be approximated as a line of zero width with a distributional
$\delta$-function energy momentum tensor. Secondly, the gravitational
field of the string is assumed to be sufficiently weak that the
linearized Einstein equations can be employed \cite{1}.
These approximations
are not valid for supermassive strings with a symmetry breaking scale
$\eta \ge m_{\rm P}$. However, for strings with $\eta \ll m_{\rm P}$,
linearized gravity is applicable almost everywhere, except in small
regions of space affected by cusps and kinks. The analysis in this
letter is based on these thin-string and weak-gravity approximations.
The metric of a static straight string lying along the $z$-axis
in cylindrical coordinates
\begin{equation}
\label{3}
ds^2=dt^2-dz^2- (1-h) (dr^2+r^2d\theta^2),
\end{equation}
where $G$ is Newton's gravitational constant, $\mu$ the string mass
per unit length and
\begin{equation}
\label{4}
h=8G \mu \ln \left( \frac{r}{r_0} \right).
\end{equation}
Introducing a new radial coordinate $r'$ as
\begin{equation}
\label{5}
(1-h) r^2=(1-8 G \mu ) r'^2,
\end{equation}
we obtain to linear order in $G\mu$,
\begin{equation}
\label{6}
ds^2=dt^2-dz^2-dr'^2-(1-8 G\mu)r'^2 d\theta^2,
\end{equation}
Finally, with a new angular coordinate
\begin{equation}
\label{7}
\theta'=(1-4G\mu)\theta,
\end{equation}
the metric takes a Minkowskian form
\begin{equation}
\label{8}
ds^2=dt^2-dz^2-dr'^2-r'^2d\theta'^2.
\end{equation}
So, the geometry around a straight cosmic string is locally identical
to that of flat spacetime. This geometry, however is not globally
Euclidean since the angle $\theta'$ varies in the range
\begin{equation}
\label{9}
0 \leq \theta' < 2 \pi (1-4G\mu).
\end{equation}
Hence, the effect of the string is to introduce an azimuthal `deficit
angle'
\begin{equation}
\label{10}
\Delta=8\pi G \mu,
\end{equation}
implying that a surface of constant $t$ and $z$ has the geometry of a
cone rather than that of a plane \cite{2}.

As shown above, the metric (\ref{6}) can be transformed to a flat metric
(\ref{8}) so there is no gravitational potential in the space outside the
string. But there is a delta-function curvature at the core of the cosmic
string which has a global effect-the deficit angle (\ref 10).

The dimensionless parameter $G \mu$ plays an important role in the
physics of cosmic strings. In the weak-field approximation $G\mu \ll 1$.
The string scenario for galaxy formation requires $G \mu \sim 10^{-6}$
while observations constrain $G \mu$ to be less than $10^{-5}$ \cite{2}.

The authors of Refs. \cite{3,4,5} have shown that the electrostatic
field of a charged particle is distorted by the cosmic string.
For a test charged particle in the presence of a cosmic string
the electrostatic self-force is repulsive with magnitude
\begin{equation}
\label{11}
F=\frac{\pi G \mu e^2}{4 r^2}.
\end{equation}
For the Bohr's atom in the absence of a cosmic string, the electrostatic
force between an electron and a proton is given by Coulomb law:
\begin{equation}
\label{12}
F=\frac{-ke^2}{r^2},
\end{equation}
where $k=1/(4\pi \epsilon_0)=8.99 \times 10^9 {\rm N.m^2/C^2}$.
The orbital speed of the electron in the first Bohr orbit is 
\begin{equation}
\label{13}
v_1=\frac{k e^2}{\hbar}.
\end{equation}
The ratio of this speed to the speed of light, $v_1/c$, is
known as the fine structute constant \cite{6}:
\begin{equation}
\label{14}
\alpha=\frac{e^2}{4 \pi \epsilon_0 \hbar c}.
\end{equation}
Inserting the known values of the constants in the right-hand side of
this equation shows that $\alpha=1/137.0360$. Thus the electron in the
first Bohr orbit moves at $\frac{1}{137}$ the speed of light.

To calculate the fine structure constant in the spacetime of a cosmic
string we consider a Bohr's atom in the presence of a cosmic string.
For a Bohr's atom in the spacetime of a cosmic string, we
take into account in Eq.(\ref{11}) the sum of two forces, i.e.
the electrostatice force for Bohr's atom in the absence of a cosmic
string, given by Eq.(\ref{12}), plus the electrostatic self-force of
the electron in the presence of a cosmic string. Therefore,
we have
\begin{equation}
\label{15}
F=-\frac{e^2}{4 \pi \epsilon_0 r^2}+\frac{\pi G \mu e^2}{4 r^2}.
\end{equation}
It can be easily shown that this force has negative value and
is an attractive force ($\pi^2 \epsilon_0 G \mu < 1$).

The numerical value of the fine structure constant in the spacetime of
a cosmic string can be computed by Eq.(\ref{15}). The orbital
speed of the electron in the first Bohr orbit in the spacetime of a
cosmic string has positive value and is given by:
\begin{equation}
\label{16}
{\hat v}_1=\frac{e^2}{4 \pi \epsilon_0 \hbar}
-\frac{\pi G \mu e^2}{4 \hbar}.
\end{equation}
The ratio of this speed to the speed of light, ${\hat v}_1/c$,
is presented by the symbol $\hat \alpha$ which is the fine structure
constant in the spacetime of a cosmic string:
\begin{equation}
\label{17}
\hat \alpha=\frac{e^2}{4 \pi \epsilon_0 \hbar c}
-\frac{\pi G \mu e^2}{4 \hbar c}.
\end{equation}
From (\ref{14}) and (\ref{17}) we obtain:
\begin{equation}
\label{18}
\frac{\alpha}{\hat \alpha}=\frac{1}{1-\pi^2 \epsilon_0 G \mu}.
\end{equation}
In the limit $G \mu \to 0$, i.e. in the absence of the cosmic string,
${\alpha}/{\hat \alpha} \to 1$.
From Eq.(\ref{18}) we obtain
$(\alpha -\hat \alpha)/\alpha= \pi^2 \epsilon_0 G \mu$.
Inserting $G\mu \simeq 10^{-6}$ and the known values of the constants
in the right-hand side of Eq.(\ref{17}) yields
\begin{equation}
\label{19}
\hat \alpha = 
\left( 1-8.736 \times 10^{-17} \right) \alpha.
\end{equation}
This means that the presence
of a cosmic string causes the value of the fine structure constant
reduces. This reduction in the value of the fine structure
constant is very small, as given in Eq.(\ref{19}).\\
{\bf Acknowledgments:}
F.N. thanks Amir and Shahrokh for useful helps.


\begin{thebibliography}{99}
\bibitem{1} A. Vilenkin, Phys. Rev. D {\bf 23} (1981) 852.
\bibitem{2} A. Vilenkin and E.P.S. Shellard, ``{\it Cosmic Strings
and other Topological Defects}'' (Cambridge University Press, 1994).
\bibitem{3} B. Linet, Phys. Rev. D {\bf 33} (1986) 1833.
\bibitem{4} A.G. Smith, in ``{\it Formation and Evolution of Cosmic
Strings}'', Eds. G.W. Gibbons, S.W. Hawking and T. Vachaspati
(Cambridge University Press, 1990).
\bibitem{5} B. Linet, Phys. Lett. {\bf A} 124 (1987) 240.
\bibitem{6} R.T. Weidner and R.L. Sells, ``{\it Elementary Modern
Physics}''"(Allyn and Bacon, 1980).
\end{thebibliography}
\end{document}